\documentclass[aps,prd,preprint,eqsecnum,floats,epsf,superscriptaddress,nofootinbib]{revtex4}
\usepackage{epsfig}
\usepackage{graphicx}
\usepackage{extarrows}
\usepackage{subfigure}

\begin{document}
\def\be{\begin{eqnarray}}
\def\en{\end{eqnarray}}
\def\non{\nonumber}
\def\ov{\overline}
\def\la{\langle}
\def\ra{\rangle}
\def\B{{\cal B}}
\def\pr{{\sl Phys. Rev.}~}
\def\prl{{\sl Phys. Rev. Lett.}~}
\def\pl{{\sl Phys. Lett.}~}
\def\np{{\sl Nucl. Phys.}~}
\def\zp{{\sl Z. Phys.}~}
\def\up{\uparrow}
\def\dw{\downarrow}
\def\lsim{ {\ \lower-1.2pt\vbox{\hbox{\rlap{$<$}\lower5pt\vbox{\hbox{$\sim$}
}}}\ } }
\def\gsim{ {\ \lower-1.2pt\vbox{\hbox{\rlap{$>$}\lower5pt\vbox{\hbox{$\sim$}
}}}\ } }

\font\el=cmbx10 scaled \magstep2{\obeylines\hfill December, 2022}

\vskip 1.5 cm

\centerline{\large\bf Doubly charmed baryon decays $\Xi_{cc}^{++}\to\Xi_c^{(\prime)+}\pi^+$ in the quark model}

\bigskip
\bigskip
\centerline{\bf Shuge Zeng, Fanrong Xu\footnote{fanrongxu@jnu.edu.cn}
}
\medskip
\centerline{Department of Physics, Jinan University}
\centerline{Guangzhou 510632, People's Republic of China}
\medskip
\medskip
\centerline{\bf Peng-Yu Niu}
\medskip
\centerline{Guangdong Provincial Key Laboratory of Nuclear Science }
\centerline{Institute of Quantum Matter, South China Normal University}
\smallskip
\centerline{and}
\smallskip
\centerline{Guangdong-Hong Kong Joint Laboratory of Quantum Matter}
\centerline{Southern Nuclear Science Computing Center, South China Normal University}
\centerline{Guangzhou 510006, People's Republic of China}
\medskip
\medskip
\centerline{\bf Hai-Yang Cheng}
\medskip
\centerline{Institute of Physics, Academia Sinica}
\centerline{Taipei, Taiwan 115, Republic of China}
\bigskip
\bigskip
\centerline{\bf Abstract}
\bigskip
\small
In this work we study the doubly charmed baryon decays $\Xi_{cc}^{++}\to\Xi_c^{(\prime)+}\pi^+$
within the framework of the non-relativistic quark model (NRQM). Factorizable amplitudes are expressed in terms of transition form factors,  while nonfactorizable amplitudes arising form the inner $W$-emission are evaluated using current algebra and the pole model and  expressed in terms of baryonic matrix elements and axial-vector form factors. Nonperturbative parameters are then calculated using the NRQM. They can be expressed in terms of the momentum integrals of baryon wave functions, which are in turn expressed in terms of the
harmonic oscillator parameters $\alpha_\rho$ and $\alpha_\lambda$ for $\rho$- and $\lambda$-mode excitation.
The measured ratio $R$ of the branching fraction of $\Xi_{cc}^{++}\to \Xi^{\prime +}_c\pi^+$ relative to $\Xi_{cc}^{++}\to \Xi_c^+\pi^+$ can be accommodated in the NRQM with  $\alpha_{\rho 1}$ and $\alpha_{\rho_2}$ being in the vicinity of 0.51 and 0.19, respectively, where $\alpha_{\rho 1}$ is the $\alpha_\rho$ parameter for $\Xi_{cc}^{++}$ and  $\alpha_{\rho 2}$ for $\Xi_{c}^{(\prime)+}$.
Decay asymmetries are predicted to be $-0.78$ and $-0.89$ for $\Xi_c^+\pi^+$ and $\Xi_c^{\prime +}\pi^+$ modes, respectively, which can be tested in the near future. We compare our results with other works and point out that although some other models can accommodate the ratio $R$, they tend to
lead to a branching fraction of $\Xi_{cc}^{++}\to \Xi_c^+\pi^+$ too large compared to that inferred from the LHCb measurement of its rate relative to $\Xi_{cc}^{++}\to\Lambda_c^+ K^- \pi^+\pi^+$.

\pagebreak

\section{Introduction}

Hadronic decays of the doubly charmed baryon $\Xi_{cc}^{++}$ have been measured through the decays $\Xi_{cc}^{++}\to \Lambda_c^+ K^-\pi^+\pi^+$ \cite{Aaij:2017ueg} and $\Xi_{cc}^{++}\to \Xi_c^+\pi^+$ \cite{Aaij:2018gfl}. Recently, the decay $\Xi_{cc}^{++}\to \Xi^{\prime +}_c\pi^+$ was first observed by LHCb \cite{LHCb:2022rpd} and its branching fraction relative to that of $\Xi_{cc}^{++}\to \Xi_c^+\pi^+$ was also reported
\begin{equation} \label{eq:R}
R\equiv \frac{\mathcal{B}(\Xi_{cc}^{++}\rightarrow\Xi_{c}^{\prime +}\pi^+)}{\mathcal{B}(\Xi_{cc}^{++}
\rightarrow\Xi_{c}^+\pi^+)}=1.41\pm0.17\pm0.10\;,
\end{equation}
while the branching fraction of $\Xi_{cc}^{++}\to \Xi_c^+\pi^+$ relative to $\Xi_{cc}^{++}\to \Lambda_c^+ K^-\pi^+\pi^+$ was measured to be \cite{Aaij:2018gfl}
\begin{equation} \label{eq:Xicpi}
\frac{\mathcal{B}(\Xi_{cc}^{++}\to\Xi_c^+\pi^+)\times \mathcal{B}(\Xi_c^+\to p K^-\pi^+)}
{\mathcal{B}(\Xi_{cc}^{++}\to\Lambda_c^+ K^- \pi^+\pi^+)\times
\mathcal{B}(\Lambda_c^+\to p K^- \pi^+)}=0.035\pm 0.009 ({\rm{stat.}})
\pm 0.003({\rm{syst.}}).
\end{equation}

\begin{figure}[h]
\begin{center}
\includegraphics[width=0.60\textwidth]{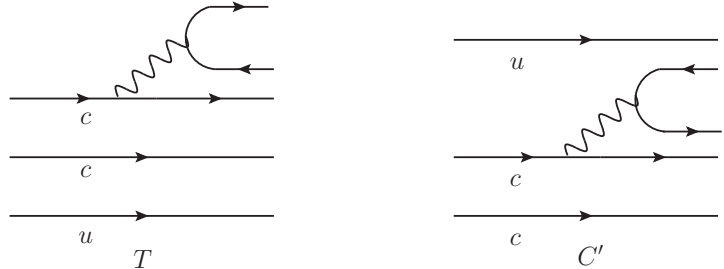}
\vspace{0.03cm}
\caption{Topological diagrams contributing to $\Xi_{cc}^{++}\to\Xi_c^{(\prime)+}\pi^+$ decays: external $W$-emission $T$ and inner $W$-emission $C'$. } \label{fig:Bicc}
\end{center}
\end{figure}

Both two-body decay modes $\Xi_c^+\pi^+$ and $\Xi^{\prime +}_c\pi^+$ proceed through the topological diagrams, external $W$-emission $T$ and inner $W$-emission $C'$ (see Fig. \ref{fig:Bicc}). \footnote[1]{Color-suppressed internal $W$-emission diagram is denoted by $C$.}
Many early studies focused only on the factorizable contribution from $T$ \cite{Wang:2017azm,Gerasimov:2019jwp,Ke:2019lcf,Shi:2019hbf}. It turns out that light-front quark model \cite{Wang:2017azm,Ke:2019lcf} and QCD sum rules \cite{Shi:2019hbf} lead to a rate of $\Xi_{cc}^{++}\to \Xi_c^+\pi^+$  larger than  that of $\Xi_{cc}^{++}\to \Xi^{\prime +}_c\pi^+$. This implies that factorizable contributions alone will  yield $R<1$. Nonfactorizable inner $W$-emission $C'$ has been considered in Refs. \cite{Gutsche:2018msz,Cheng:2020wmk,Han:2021azw,Shi:2022kfa} and partially in Ref. \cite{Sharma:2017txj}. In Ref. \cite{Han:2021azw}, nonfactorizable effects were estimated based on the final-state rescattering. The interference between $T$ and $C'$ in $\Xi_{cc}^{++}\to \Xi_c^+\pi^+$ was found to be destructive in Refs. \cite{Gutsche:2018msz,Cheng:2020wmk,Shi:2022kfa}, but constructive in Ref. \cite{Han:2021azw}. On the contrary, a large constructive interference
in the $P$-wave amplitude was obtained in Ref. \cite{Sharma:2017txj},
while nonfactorizable corrections to the $S$-wave one were not considered (see Table \ref{tab:comparison} below).

We have mentioned that factorizable contributions alone will usually  lead to $R<1$.
A possibility of accounting for the observation of $R>1$ is to consider the $\Xi_c^+- \Xi^{\prime +}_c$ mixing
\be
|\Xi_c^+\ra &=& \cos\theta\, |\Xi_c^{\bar{\bf 3}}\ra+\sin\theta\, |\Xi_c^{\bf 6}\ra, \non \\
|\Xi_c^{\prime +}\ra &=& -\sin\theta\, |\Xi_c^{\bar{\bf 3}}\ra+\cos\theta\, |\Xi_c^{\bf 6}\ra,
\en
where $\Xi_c$ and $\Xi_c^\prime$ are physical states, and $\Xi_c^{\bar{\bf 3}({\bf 6})}$ are antitriplet (sextet) charmed baryons.
As pointed out in Ref. \cite{Ke:2022gxm}, the ratio $R=0.56$ predicted in Ref. \cite{Ke:2019lcf} can be enhanced to $1.41$ when the mixing angle $\theta$ is either $16.27^\circ$ or $85.54^\circ$. However, we have to keep in mind that the effect of inner $W$-emission needs to be taken into account eventually.
Recently, the effect of $\Xi_c- \Xi^{\prime}_c$ mixing  was studied in Ref. \cite{Geng:2022yxb} in an attempt of resolving the tension between the experimental measurements and theoretical expectations in $\Xi_c^0\to \Xi_c^- e^+\nu_e$. The mixing angle was found to be $\theta=\pm 0.137(5)\pi=\pm (24.7\pm0.9)^\circ$.

Most of the studies in the literature aim at the accommodation of the ratio $R$. Although the absolute branching fractions have not been directly measured, we nevertheless can get some information on $\B(\Xi_{cc}^{++}\to \Xi_c^+\pi^+)$. Using the measurements  $\B(\Lambda_c^+\to p K^-\pi^+)=(6.28\pm0.32)\%$ and $\B(\Xi_c^+\to p K^-\pi^+)=(0.62\pm0.30)\%$ \cite{PDG2022}, it follows from Eq. (\ref{eq:Xicpi}) that
\begin{equation}
{\mathcal{B}(\Xi_{cc}^{++}\to\Xi_c^+\pi^+) \over
\mathcal{B}(\Xi_{cc}^{++}\to\Lambda_c^+ K^- \pi^+\pi^+)}=0.35\pm0.20\,.
\end{equation}
As pointed out in Ref. \cite{Cheng:2020wmk}, it is plausible to  assume that $\B(\Xi_{cc}^{++}\to\Lambda_c^+ K^- \pi^+\pi^+)\approx {2\over 3}\B(
\Xi_{cc}^{++}\to\Sigma_c^{++}\overline{K}^{*0})$. Since $\Xi_{cc}^{++}\to\Sigma_c^{++}\overline{K}^{*0}$ is a purely factorizable process, its rate can be reliably estimated once the relevant form factors are determined. Taking the latest prediction $\B(\Xi_{cc}^{++}\to\Sigma_c^{++}\overline{K}^{*0})=5.61\%$ from \cite{Gutsche:2019iac} as an example, we obtain \footnote[2]{Our previous number $(1.83\pm1.01)\%$ given in Ref. \cite{Cheng:2020wmk} is modified as the world average of the branching fraction of $\Xi_c^+\to p K^-\pi^+$ has been updated due to a new measurement from LHCb \cite{LHCb:2020gge}.}
\be \label{eq:BRexpt}
\mathcal{B}(\Xi_{cc}^{++}\to\Xi_c^+\pi^+)_{\rm expt}\approx (1.33\pm0.74)\%.
\en
Therefore, there exist two constraints: the ratio $R$ and the absolute branching fraction of
$\Xi_{cc}^{++}\to\Xi_c^+\pi^+$ inferred from the LHCb measurement of its rate relative to $\Xi_{cc}^{++}\to\Lambda_c^+ K^- \pi^+\pi^+$.

In Ref. \cite{Cheng:2020wmk} we have considered the two-body decays of doubly charmed baryons within the framework of the MIT bag model. The branching fractions of $\Xi_{cc}^{++}\to\Xi_c^+\pi^+$ and  $\Xi_{cc}^{++}\to\Xi_c^{\prime+}\pi^+$ were found to be 3.60\% and 4.65\%, respectively. At this level, $R=1.29$.
Because of a large destructive interference between $T$ and $C'$ occurred in the former mode, its branching fraction is reduced from 3.60\% to 0.69\%, whereas the latter mode is almost not affected by the internal $W$-emission owing to the Pati-Woo theorem \cite{Pati:1970fg}. Although the final branching fraction of $\Xi_{cc}^{++}\to\Xi_c^+\pi^+$ is consistent with Eq. (\ref{eq:BRexpt}), the ratio $R$ is enhanced from 1.29 to 6.74, which is evidently too large compared to experiment. Since the interference is destructive in $\Xi_c^+\pi^+$ and negligible in $\Xi_c^{\prime+}\pi^+$, this means that in order to account for the measured value of $R$, one should have $R<1$ before the inner $W$-emission is turned on.

Very recently, it was pointed out in Ref. \cite{Geng:R} that the difficulty with the bag model calculation can be overcome by considering the $\Xi_c^+-\Xi_c^{\prime+}$ mixing. At $\theta=-24.7^\circ$, one will have branching fractions 2.24\% and 3.25\%, respectively, for  $\Xi_{cc}^{++}\to\Xi_c^+\pi^+$ and  $\Xi_{cc}^{++}\to\Xi_c^{\prime+}\pi^+$. Hence $R=1.45$ is accommodated nicely and the rate of $\Xi_c^+\pi^+$ is consistent with Eq. (\ref{eq:BRexpt}).

To explore the possibility of accounting for both the ratio $R$ and the absolute branching fraction of $\Xi_{cc}^{++}\to\Xi_c^{+}\pi^+$ inferred from Eq. (\ref{eq:Xicpi}) within a phenomenological model,
in this work we shall focus on the non-relativistic quark model (NRQM) to see if we can achieve both aforementioned goals.
This paper is organized as follows. In Sec. II we follow Ref. \cite{Cheng:2020wmk} to express the factorizable and nonfactorizable amplitudes of $\Xi_{cc}^{++}\to\Xi_c^{(\prime)+}\pi^+$ decays in terms of the form factors and baryonic matrix elements which in turns are evaluated using the NRQM.  Numerical results are presented in Sec. III. We summarize our results in Section IV. Appendix A recapitulates the essences of the NRQM. Derivations of the non-perturbative parameters in the quark model are shown in Appendix B.

\section{Formalism}

The amplitude of the two-body baryonic weak decay  $\B_i\rightarrow \B_fP$ is given by
\begin{equation} \label{eq:Amp}
M(\B_i\rightarrow \B_fP)=i\bar{u}_f(A-B\gamma_5)u_i\;,
\end{equation}
where $\B_i~(\B_f)$ is the initial (final) baryon and $P$ is a pseudoscalar meson.
The decay width and up-down decay symmetry have the expressions
\begin{equation}
\begin{split}
&\Gamma=\frac{p_c}{8\pi}\left[\frac{(m_i+m_f)^2-m_P^2}{m_i^2}|A|^2+\frac{(m_i-m_f)^2-m_P^2}{m_i^2}|B|^2\right]\;,\\
&\alpha=\frac{2\kappa Re(A^*B)}{|A|^2+\kappa^2|B|^2}\;,
\end{split}
\end{equation}
with $\kappa=p_c/(E_f+m_f)$, where $p_c$ is the c.m. momentum in the rest frame of the mother baryon. The $S$- and $P$-wave amplitudes of the two-body decay generally receive both factorizable and non-factorizable contributions
\begin{equation}
A=A^{\rm fac}+A^{\rm nf}\;,\quad B=B^{\rm fac}+B^{\rm nf}\;.
\end{equation}

For doubly charmed baryon decays  $\Xi_{cc}^{++}\to\Xi_c^{(\prime)+}\pi^+$, the relevant effective Hamiltonian reads
\begin{align} \label{eq:effH}
&\mathcal{H}_{\rm{eff}}=\frac{G_F}{\sqrt{2}}V_{ud}^* V_{cs}(c_1O_1+c_2O_2)+h.c., \nonumber \\
&O_1=(\bar s c)(\bar u d),\quad
O_2=(\bar u c)(\bar s d),\qquad (\bar q_1 q_2)\equiv \bar q_1\gamma_\mu(1-\gamma_5) q_2.
\end{align}
Factorizable amplitudes read
\begin{equation} \label{eq:fac}
\begin{split}
A^{\rm fac}&=\frac{G_F}{\sqrt{2}}a_{1,2}V_{ud}^{\ast}V_{cs}f_P(m_{\B_{cc}}-m_{\B_c})f_1(q^2)\;,\\
B^{\rm fac}&=-\frac{G_F}{\sqrt{2}}a_{1,2}V_{ud}^{\ast}V_{cs}f_P(m_{\B_{cc}}+m_{\B_c})g_1(q^2)\;,
\end{split}
\end{equation}
where $f_1$ and $g_1$ are the form factors defined by
\begin{eqnarray}  \label{eq:FF}
\la \mathcal{B}_c(p_2)|\bar{c}\gamma_\mu(1-\gamma_5) u|\mathcal{B}_{cc}(p_1)\ra
=\bar{u}_2 \left[ f_1(q^2) \gamma_\mu -g_1(q^2)\gamma_\mu\gamma_5+\cdots
\right]u_1.
\end{eqnarray}
For non-factorizable contributions we follow Ref. \cite{Cheng:2020wmk} to evaluate them using current algebra and the pole model. The expressions are
\begin{equation} \label{eq:nonf}
\begin{split}
&A^{\rm nf}(\Xi_{cc}^{++}\rightarrow\Xi_{c}^+\pi^+)=\frac{1}{f_{\pi}}(-a_{\Xi_{c}^+\Xi_{cc}^{+}})\;,\\
&A^{\rm nf}(\Xi_{cc}^{++}\rightarrow\Xi_{c}^{\prime +}\pi^{+})=\frac{1}{f_{\pi}}(-a_{\Xi_{c}^{\prime +}\Xi_{cc}^{+}}),  \\
&B^{\rm nf}(\Xi_{cc}^{++}\rightarrow\Xi_{c}^+\pi^+)=\frac{1}{f_{\pi}}\left(a_{\Xi_c^+\Xi_{cc}^{+}}
\frac{m_{\Xi_{cc}^{++}}+m_{\Xi_{cc}^+}}{m_{\Xi_c^+}-m_{\Xi_{cc}^+}}g_{\Xi_{cc}^+
\Xi_{cc}^{++}}^{A(\pi^+)}\right)\;,\\
&B^{\rm nf}(\Xi_{cc}^{++}\rightarrow\Xi_{c}^{\prime+}\pi^+)=\frac{1}{f_{\pi}}\left(a_{\Xi_c^{\prime+}
\Xi_{cc}^{+}}\frac{m_{\Xi_{cc}^{++}}+m_{\Xi_{cc}^+}}{m_{\Xi_c^{\prime+}}-m_{\Xi_{cc}^+}}
g_{\Xi_{cc}^+\Xi_{cc}^{++}}^{A(\pi^+)}\right)\;,
\end{split}
\end{equation}
where $a_{\B_{f}\B_{i}}\equiv\langle  \B_f|{\cal H}_{\rm eff}^{\rm PC}| \B_i\rangle$ are baryonic matrix elements with
${\cal H}_{\rm eff}^{\rm PC}$ being the parity-conserving part of the effective Hamiltonian
and $g_{\B'\B}^{A(P)}$ are axial-vector form factors. The matrix element can be recast to the form
\begin{equation}
a_{\B_{f}\B_{i}}
=\frac{G_F}{2\sqrt{2}}V_{ud}^*V_{cs}\,c_-\langle  \B_f|O_-^{\rm PC}| \B_i\rangle
\;,
\end{equation}
where $c_\pm=c_1\pm c_2$ and $O_\pm=(\bar{s}c)(\bar{u}d)\pm(\bar{s}d)(\bar{u}c)$.

\begin{table}[t]
  \begin{center}
  \caption{Non-perturbative parameters relevant for $\Xi_{cc}^{++}\to\Xi_c^{(\prime)+}\pi^+$ decays in the NRQM.}
     \label{tab:ta1}
  \vskip 0.2cm
    \begin{tabular}{c c c}
    \hline\hline
    $ $& ~~$\Xi_{cc}^{++}\rightarrow\Xi_{c}^+\pi^+$~~ & ~~$\Xi_{cc}^{++}\rightarrow\Xi_{c}^{\prime+}\pi^+$~~\\
    \hline
    $f_1$&$\frac{\sqrt{6}}{2}X$&$\frac{\sqrt{2}}{2}X$\\
    $g_1$&$\frac{1}{\sqrt{6}}X$&$\frac{5\sqrt{2}}{6}X$\\
    $g_{\Xi_{cc}^+\Xi_{cc}^{++}}^{A(\pi^+)}$&$-\frac{1}{3}Y$&$-\frac{1}{3}Y$\\
    $\langle  \B_f|O_-| \B_i\rangle$ & $4\sqrt{6}Z$ & $0$\\
    \hline
    \hline
    \end{tabular}
  \end{center}
\end{table}

In the NRQM, the non-perturbative parameters $f_1$, $g_1$, $g_{\B'\B}^{A(P)}$ and  $\langle  \B_f|O_-| \B_i\rangle$ can be expressed in terms of the
momentum integrals of baryon wave functions  $X$, $Y$ and $Z$ given in Eq. (\ref{eq:X,Y,Z})
(see Appendix B for details)
\begin{equation}
\begin{split}
f_1 &=\langle \B_f\uparrow|b_{u}^{\dagger}b_{c}|\B_{i}\uparrow\rangle X\;, \\ g_1 &=\langle \B_f\uparrow|b_{u}^{\dagger}b_{c}\sigma_z|\B_{i}\uparrow\rangle X\;, \\
g_{\B'\B}^{A(P)} &=\langle \B_f\uparrow|b_{d}^{\dagger}b_{u}\sigma_z|\B_i\uparrow\rangle Y\;, \\
\langle \mathcal{B}_f|(\bar{q}_{1}{q}_{2})(\bar{q}_{3}{q}_{4})|\mathcal{B}_i\rangle
& =
\langle \mathcal{B}_f\uparrow|
({b}^{\dagger}_{q_1}{b}_{q_2})_1({b}^{\dagger}_{q_3}{b}_{q_4})_2
(1-\boldsymbol{\sigma}_1\cdot\boldsymbol{\sigma}_2)|\mathcal{B}_i\uparrow\rangle Z
\;,
\end{split}\label{Section2Nonfactorizable}
\end{equation}
where the subscripts 1 and 2 appearing in the last line indicate that the quark operator acts only on the first and second quarks, respectively, in the
baryon wave function.
The coefficients $\langle \B_f\uparrow|\cdots|\B_{i}\uparrow\rangle$ depend on the spin-flavor functions of baryons and they are displayed in Table \ref{tab:ta1}.
The momentum integrals can be expressed in terms of the harmonic oscillator parameters  $\alpha_\rho$ and $\alpha_\lambda$ for $\rho$- and $\lambda$-mode excitation, respectively. We shall use $\alpha_{\rho 1}$, $\alpha_{\lambda 1}$  for $\Xi_{cc}^{++}$, $\alpha_{\rho 2}$, $\alpha_{\lambda 2}$ for both $\Xi_{c}^+$ and $\Xi_{c}^{'+}$ and $\alpha_{\rho 3}$, $\alpha_{\lambda 3}$ for $\Xi_{cc}^+$.
Explicitly (see Appendix B),
\begin{equation}
\begin{split}
X&=\left(\frac{16(m_s+m_u)^2\alpha_{\lambda 1}\alpha_{\lambda 2}\alpha_{\rho 1}\alpha_{\rho 2}}{D_1+D_2}\right)^{3/2}, \\
Y&=8\left
(\frac{\alpha_{\lambda 1} \alpha_{\lambda 3} \alpha_{\rho 1}\alpha_{\rho 3}}
{(\alpha_{\lambda 1}^2 +\alpha_{\lambda 3}^2)
(\alpha_{\rho 1}^2 +\alpha_{\rho 3}^2)}\right)^{3/2}\;,\\
Z&=128\sqrt{2}\pi^{3/2}\left[\frac{\alpha_{\lambda 2}\alpha_{\lambda 3}\alpha_{\rho 2}\alpha_{\rho 3}}{4\alpha_{\lambda 2}^2+\alpha_{\lambda 3}^2+4\alpha_{\rho 3}^2}\right]^{3/2}\;,\\
\end{split}\label{eq:XYZ}
\end{equation}
where
\begin{equation}
\begin{split}
 D_1 &= (m_s+m_u)^2[4\alpha_{\rho 1}^2\alpha_{\rho 2}^2+\alpha_{\lambda 1}^2(\alpha_{\rho 1}^2+4\alpha_{\rho 2}^2)], \\
 D_2 &= \alpha_{\lambda 2}^2\left[(2m_s+m_u)^2\alpha_{\lambda 1}^2+4\left(m_u^2\alpha_{\rho 1}^2+[m_s+m_u]^2\alpha_{\rho 2}^2\right)\right],
\end{split}
\end{equation}
and
\begin{equation}
\begin{split}
\alpha_{\lambda 1} &=\left[\frac{16m_u}{3(2m_c+m_u)}\right]^{\frac{1}{4}}\alpha_{\rho 1}\;,\quad
\alpha_{\lambda 3}=\left[\frac{16m_d}{3(2m_c+m_d)}\right]^{\frac{1}{4}}\alpha_{\rho 3}\;, \\
\alpha_{\lambda 2} &=\left[\frac{4m_c(m_s+m_u)^2}{3m_sm_u(m_s+m_u+m_c)}\right]^{\frac{1}{4}}\alpha_{\rho 2}\;.
\end{split}
\end{equation}
Thus only $\alpha_{\rho 1}$, $\alpha_{\rho 2}$ and $\alpha_{\rho 3}$ are independent.

\section{Numerical results}
In the NRQM we shall take the parameters as follows:  $m_s=0.45$ GeV, $m_u=m_d=0.33$ GeV for light quark masses \cite{Arifi:2021} and $m_c=1.6$ GeV for the charm quark mass.
The parameter $\alpha_{\rho 1}$ is a harmonic oscillator parameter in the spatial wave function of the $\rho$-mode excitation between the two charm quarks. It has been taken to be $\alpha_{\rho 1}=0.47$ GeV as in the charmonium system \cite{Xiao:2017udy}. \footnote[3]{It should be stressed that a set of the Jacobi coordinate defined in Eq. (\ref{eq:2ndJacobi}) has been adopted
in Ref. \cite{Xiao:2017udy} so that the harmonic oscillator parameters $\tilde{\alpha}_\rho$ and $\tilde{\alpha}_\lambda$
respect the relation $\tilde{\alpha}_\lambda=[3m_q/(2m_Q+m_q)]^{1/4}\tilde{\alpha}_\rho$. The tilde and untilde harmonic oscillator parameters are related through Eq. (\ref{eq:primeunprime}). We have translated the result of $\tilde{\alpha}_{\rho 1}=0.66$ GeV in Ref. \cite{Xiao:2017udy} into
$\alpha_{\rho 1}=0.47$ GeV.}
The parameter $\alpha_{\rho 2}$ was determined to be 0.25 GeV in Ref. \cite{Niu:2021qcc}.
Notice that for $\Lambda_c^+$, $\alpha_\rho$ ranges from 0.26 to 0.32 GeV \cite{Nagahiro:2016nsx,Arifi:2021,Niu:2020}.
For $\alpha_{\rho 3}$, we shall take $\alpha_{\rho 3}=\alpha_{\rho 1}$ as it should be the same as $\alpha_{\rho 1}$ in the isospin limit.

\begin{table}[t]
\center
\caption{Form factors $f_1, g_1$, the axial-vector form factor $g_{\B'\B}^{A(\pi)}$ and baryonic matrix elements $\langle  \B_f|O_-| \B_i\rangle$ calculated in the NRQM with the specified harmonic oscillator parameters $(\alpha_{\rho 1},\alpha_{\rho 2})$ in units of GeV. The results of Ref. \cite{Cheng:2020wmk} obtained from the bag model are also shown here for comparison.
} \label{tab:FFs}
\vspace{6pt}
\setlength{\tabcolsep}{1.7mm}
\footnotesize{
{
\begin{tabular}{c l c llrl}
\hline\hline
\multicolumn{2}{c}{}                   &   $(\alpha_{\rho 1}, \alpha_{\rho 2})$ & $f_1(m_P^2)$   &  $g_1(m_P^2)$   & $g_{\B'\B}^{A(\pi)}$  & $\langle  \B_f|O_-| \B_i\rangle$    \\ \hline
\multicolumn{2}{l}{$\Xi_{cc}^{++}\rightarrow\Xi_{c}^{+}\pi^+$}                   &          &         \\
\multicolumn{2}{l}{Case 1}                   &   (0.50, 0.21)        &    0.709 &   0.236 & $-$0.333 & ~~0.0310 \\
\multicolumn{2}{l}{Case 2}                   &    (0.51, 0.19)        &    0.574 &   0.191 & $-$0.333 & ~~0.0247    \\
\multicolumn{2}{l}{Case 3}                   &   (0.53, 0.17)        &    0.425 &   0.141 & $-$0.333 & ~~0.0191 \\
\multicolumn{2}{l}{CMXZ \cite{Cheng:2020wmk}}                 &       &    0.577 &   0.222 & $-$0.217 & ~~0.0214    \\ \hline
\multicolumn{2}{l}{$\Xi_{cc}^{++}\rightarrow\Xi_{c}^{\prime +}\pi^+$}                   &          &         \\
\multicolumn{2}{l}{Case 1}                   &   (0.50, 0.21)        &    0.397 &   0.662 & $-$0.333 & ~~~~~0         \\
\multicolumn{2}{l}{Case 2}                   &    (0.51, 0.19)        &    0.323 &   0.538 & $-$0.333 & ~~~~~0     \\
\multicolumn{2}{l}{Case 3}                   &   (0.53, 0.17)        &    0.240 &   0.400 & $-$0.333 & ~~~~~0    \\
\multicolumn{2}{l}{CMXZ \cite{Cheng:2020wmk}}                  &      & 0.386    &   0.703 &  $-$0.217  & ~~$8.4\times10^{-5}$     \\ \hline\hline
\end{tabular}}}
\end{table}

Form factors $f_1$ and  $g_1$, the axial-vector form factor $g_{\B'\B}^{A(\pi)}$ and baryonic matrix elements $\langle  \B_f|O_-| \B_i\rangle$ are calculated in the NRQM using Eq. (\ref{Section2Nonfactorizable}) and Table \ref{tab:ta1}. The numerical results are exhibited in Table \ref{tab:FFs}
with several specified harmonic oscillator parameters $(\alpha_{\rho 1},\alpha_{\rho 2})$ to be discussed below. We also show the bag model results obtained in Ref. \cite{Cheng:2020wmk} for comparison. Notice that the matrix element for $\Xi_{cc}^{++}\to \Xi_c^{\prime +}$ transition receives contributions only from the small component of the quark wave function in the bag model and hence it vanishes in the NRQM. In the bag model the matrix element for $\Xi_{cc}^{++}\to \Xi_c^{\prime +}$ transition is nonzero, but it is quite suppressed relative to the matrix element $a_{\Xi_c^+ \Xi_{cc}^{++}}$.

\begin{figure}[t]
\begin{center}
\includegraphics[width=0.60\textwidth]{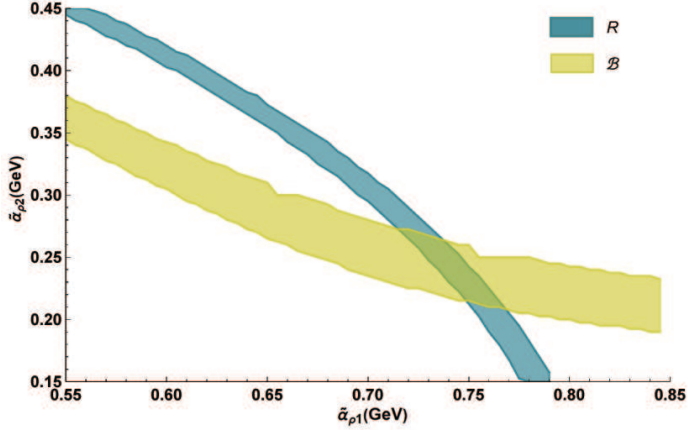}
\vspace{0.03cm}
\caption{Allowed regions for the harmonic oscillator parameters $\alpha_{\rho 1}$ and
$\alpha_{\rho 2}$ constrained by the ratio $R$ and the absolute branching fraction of
$\Xi_{cc}^{++}\to\Xi_c^+\pi^+$ inferred from  Eq. (\ref{eq:BRexpt}). }
\label{fig:alpha}
\end{center}
\end{figure}

We plot in Fig. \ref{fig:alpha} the allowed regions for the harmonic oscillator parameters $\alpha_{\rho 1}$ and $\alpha_{\rho 2}$ constrained by the ratio of branching fractions $R$ [see Eq. (\ref{eq:R})] and the absolute branching fraction of
$\Xi_{cc}^{++}\to\Xi_c^+\pi^+$ inferred from Eq. (\ref{eq:BRexpt}). It is clear that the allowed range of $0.505\sim 0.545$ GeV for $\alpha_{\rho 1}$ is compatible with the value of 0.47 GeV inferred from the charmonium system. However,  the preferred range of $\alpha_{\rho 2}$, $(0.145\sim 0.195)$ GeV, is somewhat smaller than the naive expectation of 0.25 GeV.
Accordingly, in Table \ref{tab:FFs} we choose three sets of harmonic oscillator parameters denoted by cases 1, 2 and 3 with $(\alpha_{\rho 1},\alpha_{\rho 2})$ being in the vicinity of 0.51 and 0.19 GeV, respectively.

\begin{table}[p!]
\center
\caption{Comparison of the predicted $S$- and $P$-wave amplitudes (in units of $10^{-2}G_F$ GeV$^2$) of $\Xi_{cc}^{++}\to\Xi_c^{(\prime)+}\pi^+$ decays, their branching fractions (in units of $10^{-2}$) and the decay asymmetry parameter $\alpha$ in various approaches with only the central values being cited.   For the predictions of Sharma and Dhir \cite{Sharma:2017txj}, we quote only the flavor-independent pole amplitudes for both NRQM and HQET. Two bag models are considered in the work of Liu and Geng \cite{Geng:R}: the static bag (SB) and homogeneous bag (HB) models.
}
\setlength{\tabcolsep}{2.5mm}
\resizebox{\textwidth}{!} {
{
\begin{tabular}{c r r r r r r c r r c}
\hline\hline
\multicolumn{2}{c}{}    &  $A^{\rm fac}$  & $A^{\rm nf}$ & $A^{\rm tot}$ & $B^{\rm fac}$ &  $B^{\rm nf}$  & \multicolumn{1}{r|}{$B^{\rm tot}$ }  & $\mathcal{B}$  &    \multicolumn{1}{c|}{$ \alpha$} & $R$ \\ \hline
\multicolumn{2}{l}{$\Xi_{cc}^{++}\rightarrow\Xi_{c}^{+}\pi^+$}                   &       &         &         &        &    & \multicolumn{1}{c|}{}   &    &  \multicolumn{1}{c|}{} &  \\
\multicolumn{2}{l}{This work}      &   &    &   &  & & \multicolumn{1}{c|}{}   &  &   \multicolumn{1}{c|}{} & \\
\multicolumn{2}{l}{\;\;\; Case 1}                   &  9.1   & $-$15.6   &  $-$6.5  & $-$16.0  & 27.4 & \multicolumn{1}{r|} {11.4}   & 3.01    & \multicolumn{1}{c|}{ $-$0.78}&  \\
\multicolumn{2}{l}{\;\;\; Case 2}                  &  7.4   & $-$12.4   &  $-$5.0  & $-$13.0  &21.8 & \multicolumn{1}{r|}{8.8}   & 1.83    & \multicolumn{1}{c|}{ $-$0.78}& \\
\multicolumn{2}{l}{\;\;\; Case 3}                  &  5.5   & $-$9.6   &  $-$4.1  & $-$9.6  &16.8 & \multicolumn{1}{r|}{7.2}   & 1.20    & \multicolumn{1}{c|}{ $-$0.78}& \\
\multicolumn{2}{l}{CMXZ \cite{Cheng:2020wmk}}   &  7.4  &   $-$10.8 &  $-$3.4  &  $-$15.1 &18.9 & \multicolumn{1}{r|}{3.8}   & 0.69 &   \multicolumn{1}{c|}{ $-$0.41 } & \\
\multicolumn{2}{l}{Gutsche {\it et al.} \cite{Gutsche:2018msz}}   &  $-$8.1  &  11.5 &  3.4  &  13.0 & $-$18.5 & \multicolumn{1}{r|}{$-$5.6}   & 0.71 &   \multicolumn{1}{c|}{ $-$0.57 } & \\
\multicolumn{2}{l}{Sharma\;\&\;Dhir \cite{Sharma:2017txj}}      &   &    &   &  & & \multicolumn{1}{c|}{}   &  &   \multicolumn{1}{c|}{} & \\
\multicolumn{2}{l}{\;\;\; NRQM}      &  7.38 & 0   & 7.38  &  $-$16.77 & $-$24.95 & \multicolumn{1}{r|}{$-$41.72}   & 6.64 &    \multicolumn{1}{c|}{$-$0.99} &  \\
\multicolumn{2}{l}{\;\;\; HQET}    &  9.52 & 0   & 9.52  &  $-$19.45 & $-$24.95 & \multicolumn{1}{c|}{$-$44.40}   & 9.19 &  \multicolumn{1}{c|}{$-$0.99 } & \\
\multicolumn{2}{l}{Shi {\it et al.} \cite{Shi:2022kfa}}       &   &    & &   & & \multicolumn{1}{c|}{}   &  &   \multicolumn{1}{c|}{} & \\
\multicolumn{2}{l}{\;\;\; LCSR+HQET}               &  9.52 & $-$16.67  & $-$7.18  &  $-$19.45 & $-$20.47 & \multicolumn{1}{r|}{$-$39.92}   & 6.22 &  \multicolumn{1}{c|}{$+$0.99} & \\
\multicolumn{2}{l}{Ke \& Li \cite{Ke:2022gxm}}      &   &    &   &  & & \multicolumn{1}{c|}{}   &  &
\multicolumn{1}{c|}{} & \\
\multicolumn{2}{l}{\;\;\; $\theta=16.27^\circ$}      &   &    &   &  &  & \multicolumn{1}{c|}{}   & 2.14 &   \multicolumn{1}{c|}{$-$0.09}&  \\
\multicolumn{2}{l}{\;\;\; $\theta=85.54^\circ$}      &   &    &   &  &  & \multicolumn{1}{c|}{}   & 2.14 &   \multicolumn{1}{c|}{$-$0.95}&  \\
\multicolumn{2}{l}{Liu\;\&\;Geng \cite{Geng:R}\footnote[4]{We wish to thank C. W. Liu and C. Q. Geng for providing us the numerical values of the $S$- and $P$-wave amplitudes in their work.} }      &   &    &   &  & & \multicolumn{1}{c|}{}   &  &   \multicolumn{1}{c|}{} & \\
\multicolumn{2}{l}{\;\;\; SB ($\theta=-24.7^\circ$)}      & $4.83$ & $-9.99$ & $-5.16$ &$5.16$ &$13.6$ & \multicolumn{1}{r|}{18.8}   & 2.24 &    \multicolumn{1}{c|}{$-$0.93} & \\
\multicolumn{2}{l}{\;\;\; HB ($\theta=24.7^\circ$)}      & 7.08  & $-20.3$   & $-13.2$  & $-22.1$ &  33.0 &  \multicolumn{1}{r|}{10.9}   & 10.3 &  \multicolumn{1}{c|}{$-$0.30 } & \\
\hline
\multicolumn{2}{l}{$\Xi_{cc}^{++}\rightarrow\Xi_{c}^{\prime +}\pi^+$}    &          &         &         &        &    & \multicolumn{1}{c|}{}   &    &   \multicolumn{1}{c|}{}& \\
\multicolumn{2}{l}{This work}      &   &    &   &  & & \multicolumn{1}{c|}{}   &  &   \multicolumn{1}{c|}{} & \\
\multicolumn{2}{l}{\;\;\; Case 1}     &   4.6  &  0  & 4.6   & $-$45.6  &  0 & \multicolumn{1}{r|}{$-$45.6}   & 4.32   & \multicolumn{1}{c|}{$-$0.89 } & 1.44 \\
\multicolumn{2}{l}{\;\;\; Case 2}     &   3.7  &  0  & 3.7   & $-$37.1  &  0 & \multicolumn{1}{r|}{$-$31.0}   & 2.86   & \multicolumn{1}{c|}{$-$0.89 }& 1.56 \\
\multicolumn{2}{l}{\;\;\; Case 3}     &   2.8  &  0  & 2.8   & $-$27.6  &  0 & \multicolumn{1}{r|}{$-$27.6}   & 2.16   & \multicolumn{1}{c|}{$-$0.89 }& 1.32 \\
\multicolumn{2}{l}{CMXZ \cite{Cheng:2020wmk}}          & 4.5    & $-$0.04  &   4.5 & $-$48.5  & $-$0.06   & \multicolumn{1}{r|}{$-$48.4}   &  4.65  & \multicolumn{1}{c|}{ $-$0.84 }& 6.74 \\
\multicolumn{2}{l}{Gutsche {\it et al.} \cite{Gutsche:2018msz}}   &  $-$4.3  &  $-$0.1 & $-$4.4  & 37.6 & 1.4 & \multicolumn{1}{r|}{39.0}   & 3.39 &   \multicolumn{1}{c|}{ $-$0.93 } & 4.33 \\
\multicolumn{2}{l}{Sharma\;\&\;Dhir \cite{Sharma:2017txj}}   &   &    &   &  & & \multicolumn{1}{c|}{}   &  &  \multicolumn{1}{c|}{} & \\
\multicolumn{2}{l}{\;\;\; NRQM}             &  4.29 & 0   & 4.29  &  $-$53.65 & 0 & \multicolumn{1}{r|}{$-$53.65}   & 5.39 &   \multicolumn{1}{c|}{$-$0.78 }&  0.81\\
\multicolumn{2}{l}{\;\;\; HQET}           &  5.10 & 0   & 5.10  &  $-$62.37 &0 & \multicolumn{1}{r|}{$-$62.37}   & 7.34 &   \multicolumn{1}{c|}{$-$0.79} &  0.80 \\
\multicolumn{2}{l}{Shi {\it et al.} \cite{Shi:2022kfa}}            &   &    & &   & & \multicolumn{1}{c|}{}   &  &   \multicolumn{1}{c|}{} & \\
\multicolumn{2}{l}{\;\;\; LCSR+HQET}  &  5.10 & $-$0.83   & 4.27  &  $-$62.37 & $-$8.86 & \multicolumn{1}{r|}{$-$71.23}   & 8.85 &   \multicolumn{1}{c|}{$-$0.64} & 1.42\\
\multicolumn{2}{l}{Ke \& Li \cite{Ke:2022gxm}}      &   &    &   &  & & \multicolumn{1}{c|}{}   &  &
\multicolumn{1}{c|}{} & \\
\multicolumn{2}{l}{\;\;\; $\theta=16.27^\circ$}      &   &    &   &  &  & \multicolumn{1}{c|}{}   & 3.02 &   \multicolumn{1}{c|}{$-$0.99 } &  1.41 \\
\multicolumn{2}{l}{\;\;\; $\theta=85.54^\circ$}      &   &    &   &  &  & \multicolumn{1}{c|}{}   & 3.02 &   \multicolumn{1}{c|}{$-$0.51} & 1.41 \\
\multicolumn{2}{l}{Liu\;\&\;Geng \cite{Geng:R}}      &   &    &   &  & & \multicolumn{1}{c|}{}   &  &   \multicolumn{1}{c|}{} & \\
\multicolumn{2}{l}{\;\;\; SB ($\theta=-24.7^\circ$)}      &  7.38 & $-4.82$   & 2.56  & $-51.0$ & 7.26 & \multicolumn{1}{r|}{$-43.7$}   & 3.25 &    \multicolumn{1}{c|}{$-$0.63} & 1.45 \\
\multicolumn{2}{l}{\;\;\; HB ($\theta=24.7^\circ$)}     &  0.61  & 9.65   & 10.3  & $-28.1$ & $-17.4$&  \multicolumn{1}{r|}{$-45.5$}   & 8.91 &  \multicolumn{1}{c|}{$-$0.96 } & 0.87 \\
\hline
\multicolumn{2}{l}{LHCb \cite{LHCb:2022rpd}}           &  &  & &   & & \multicolumn{1}{c}{}   &     & \multicolumn{1}{c|}{} &$1.41\pm0.20$ \\
\hline\hline
\end{tabular}
 \label{tab:comparison}
}}
\\
\end{table}


With the input for various parameters from Table \ref{tab:FFs} we are ready to compute the factorizable and nonfactorizable amplitudes for both $S$- and $P$-waves using Eqs. (\ref{eq:fac}) and (\ref{eq:nonf}). The numerical results of individual $S$- and $P$-wave amplitudes, branching fractions of $\Xi_{cc}^{++}\to\Xi_c^{(\prime)+}\pi^+$ decays and their decay asymmetries are shown in Table \ref{tab:comparison} for three different sets of the harmonic oscillator parameters $\alpha_{\rho 1}$ and $\alpha_{\rho 2}$ given in Table \ref{tab:FFs}. Evidently, the interference between the factorizable diagram $T$ and the nonfactorizable $C'$ is destructive in
$\Xi_{cc}^{++}\to\Xi_c^{+}\pi^+$.  On the contrary, the decay $\Xi_{cc}^{++}\to\Xi_c^{\prime +}\pi^+$ does not receive nonfactorizable contributions. This is consistent with the so-called  Pati-Woo theorem \cite{Pati:1970fg} which results from the facts that the $(V-A)\times(V-A)$ structure of weak interactions is invariant under the Fierz transformation and that the baryon wave function is color antisymmetric. As a consequence of this theorem,  the quark
pair in a baryon produced by weak interactions be antisymmetric in flavor. Since the sextet
$\Xi'_c$ is symmetric in light quark flavor, it cannot contribute to $C'$. Because the form factor $g_1$ in the $\Xi_{cc}^{++}\to \Xi_c^{\prime+} \pi^+$ mode is larger than that in $\Xi_c^+ \pi^+$ (see Table \ref{tab:FFs}), the $P$-wave amplitude of the former is much higher than that of the latter. Consequently, the branching fraction of the $\Xi_c^{\prime+} \pi^+$ mode is larger than the $\Xi_c^+ \pi^+$ one.
Taking case 2 as an example, we have the results
\begin{equation}
\begin{split}
&\mathcal{B}(\Xi_{cc}^{++}\rightarrow\Xi_{c}^{+}\pi^+)=1.83\%\;, \qquad \alpha(\Xi_{c}^{+}\pi^+)=-0.78\,, \\
&\mathcal{B}(\Xi_{cc}^{++}\rightarrow\Xi_{c}^{'+}\pi^+)=2.86\%\;, \qquad
\alpha(\Xi_{c}^{\prime +}\pi^+)=-0.89\,,
\end{split}
\end{equation}
and hence $R=1.56$\,.

In Table \ref{tab:comparison} we also compare our results with other approaches.  The nonfactorizable effects have been evaluated in two entirely different approaches: current algebra and the pole model in Ref. \cite{Cheng:2020wmk} and the covariant confined quark model in Ref. \cite{Gutsche:2018msz}. It is interesting to notice that both approaches yielded a large destructive interference in  $\Xi_{cc}^{++}\rightarrow\Xi_{c}^{+}\pi^+$ and obtained similar branching fractions of order 0.70\%. Although this is consistent with the experimental value of Eq. (\ref{eq:BRexpt}) to the lower end, the predicted ratios 6.74 in  Ref. \cite{Cheng:2020wmk} and 4.33 in Ref. \cite{Gutsche:2018msz} are too large compared to the LHCb value of $1.41\pm0.20$ \cite{LHCb:2022rpd}. In the work of Sharma and Dhir \cite{Sharma:2017txj},  a large constructive interference
in the $P$-wave amplitude was found, while nonfactorizable corrections to the $S$-wave one were not considered.  From Table \ref{tab:comparison} we see that the $P$-wave amplitude in this model is much larger than other works.
$\B(\Xi_{cc}^{++}\rightarrow\Xi_{c}^{+}\pi^+)$ of order $(7-9)\%$ ($(13-16)\%$) for flavor-independent (flavor-dependent) pole amplitudes was obtained in this work, which is obviously too large compared to Eq. (\ref{eq:BRexpt}).

In the recent work of Shi {\it et al.} \cite{Shi:2022kfa}, nonfactorizable internal $W$-emission
contributions to $\Xi_{cc}^{++}\to\Xi_c^{(\prime)+}\pi^+$ decays were evaluated using light-cone sum rules, see $A^{\rm nf}$ and $B^{\rm nf}$ terms shown in Table  \ref{tab:comparison}. Factorizable contributions were then taken from the work of Sharma and Dhir \cite{Sharma:2017txj} under ``HQET".  The sizable nonfactorizable contribution to the $P$-wave of $\Xi_{cc}^{++}\to\Xi_c^{\prime +}\pi^+$ seems to be an issue in view of the Pati-Woo theorem.
From Table  \ref{tab:comparison} we see that  the $S$-wave amplitude denoted by $A^{\rm tot}$ for $\Xi_{cc}^{++}\to\Xi_c^{+}\pi^+$ is modified from 9.52 to $-7.18$ owing to the presence of a destructive nonfactorizable contribution. \footnote[5]{Recall that the relative sign convention between $S$- and $P$-waves is defined in Eq. (\ref{eq:Amp}).}
Consequently, $S$- and $P$-wave amplitudes are of the same sign in this model and yield a {\it positive} decay asymmetry $\alpha=0.99$, in sharp contrast to the other works where the decay asymmetry is always predicted to be negative. Hence, even a sign measurement of $\alpha(\Xi_c^{+}\pi^+)$ will allow to discriminate the model of Shi {\it et al.} from others.

As discussed in the Introduction, the external $W$-emission diagram $T$ alone usually  leads to $R<1$. It was first pointed out by Ke and Li \cite{Ke:2022gxm} that  the observation of $R>1$ can be accommodated by considering the $\Xi_c^+-
\Xi^{\prime +}_c$ mixing. Two mixing angles were found, $\theta=16.27^\circ$ or $85.54^\circ$ (see Table \ref{tab:comparison}). However, when the nonfactorizable effect due to internal $W$-emission is turned on, the mixing angle will be affected.  As noticed in passing, when the  $\Xi_c^+- \Xi^{\prime +}_c$ mixing effect is applied to the static bag model calculation performed
in Ref. \cite{Cheng:2020wmk}, Liu ad Geng \cite{Geng:R} have shown that at the mixing angle $\theta=-0.137\pi$, the ratio $R$ is well accommodated and the branching fraction  of $\Xi_{cc}^{++}\to\Xi_c^{+}\pi^+$
is consistent with the constraint derived from Eq. (\ref{eq:BRexpt}).

However, there is one issue with the static bag model, namely, a static bag is not invariant under space translation and it is impossible for a static bag to be at rest. The unwanted center-of-mass motion (CMM) of the bag model is an issue and it should be removed for a consistent treatment \cite{Liu:2022pdk}. For example, the bag model calculation for the heavy-flavor-conserving decays are improved by removing CMM corrections. The predictions for  $\Xi_c^0\to\Lambda_c^+\pi^-$ and $\Xi_b^-\to \Lambda_b^0\pi^-$ are both in good agreement with experiment \cite{Cheng:2022jbr}. It is clear from Table  \ref{tab:comparison} that nonfactorizable $S$- and $P$-wave amplitudes of $\Xi_{cc}^{++}\to\Xi_c^{\prime +}\pi^+$ are no longer subject to the constraint from the Pati-Woo theorem because of the contribution from $\Xi_{cc}^{++}\to
(\Xi_c^{\bar{\bf 3}})^+\pi^+$.
Unfortunately, the same bag model without CMM will lead to an even smaller ratio, $R=0.19$ \cite{Geng:R}. When the $\Xi_c^+-\Xi_c^{\prime+}$ mixing is included, $R$ is increased to 0.90 at $\theta=24.7^\circ$, but the branching fraction of $\Xi_{cc}^{++}\to\Xi_c^{\prime+}\pi^+$ becomes 10.3\% which is too large compared to the constraint inferred from Eq. (\ref{eq:BRexpt}).

\section{Conclusions}
In this work we have studied the doubly charmed baryon decays $\Xi_{cc}^{++}\to\Xi_c^{(\prime)+}\pi^+$
within the framework of the NRQM. Factorizable amplitudes are expressed in terms of transition form factors,  while nonfactorizable amplitudes arising form the inner $W$-emission are evaluated using current algebra and the pole model and expressed in terms of baryonic matrix elements and axial-vector form factors.

We draw some conclusions from our analysis:
\begin{itemize}

\item
Nonperturbative parameters are calculated in the NRQM. They can be expressed in terms of the momentum integrals of baryon wave functions, which are in turns expressed in terms of the
harmonic oscillator parameters $\alpha_\rho$ and $\alpha_\lambda$ for $\rho$- and $\lambda$-mode excitation, respectively.

\item Denoting the harmonic oscillator parameters $\alpha_{\rho 1}$, $\alpha_{\lambda 1}$  for $\Xi_{cc}^{++}$ and $\Xi_{cc}^{+}$, $\alpha_{\rho 2}$, $\alpha_{\lambda 2}$ for $\Xi_{c}^+$ and $\Xi_{c}^{'+}$, we found that the measured ratio $R$ of the branching fraction of $\Xi_{cc}^{++}\to \Xi^{\prime +}_c\pi^+$ relative to $\Xi_{cc}^{++}\to \Xi_c^+\pi^+$
can be accommodated in the NRQM with $\alpha_{\rho 1}$ and $\alpha_{\rho_2}$ being in the vicinity of 0.51 and 0.19 GeV, respectively.

\item
We have compared our results with other approaches. While the ratio $R$ has been accommodated in some other models, the predicted branching fraction of $\Xi_{cc}^{++}\to \Xi_c^+\pi^+$ is often too large compared to that inferred from the LHCb measurement of its rate relative to $\Xi_{cc}^{++}\to\Lambda_c^+ K^- \pi^+\pi^+$.

\item
Decay asymmetries are predicted to be $-0.78$ and $-0.89$ for $\Xi_c^+\pi^+$ and $\Xi_c^{\prime +}\pi^+$ modes, respectively, which can be tested in the near future.

\item
Although the static bag model fails to account for the ratio $R$, it is interesting to notice that when  the $\Xi_c^+- \Xi^{\prime +}_c$ mixing effect is taken into account in the bag model
calculations, data can be nicely accommodated with the mixing angle $\theta=-24.7^\circ$.

\end{itemize}

	\begin{acknowledgments}
		We would like to thank Chia-Wei Liu for valuable discussions and for providing us the numerical values of the $S$- and $P$-wave amplitudes presented in Ref. \cite{Geng:R}.
		This research was supported in part by the Ministry of Science and Technology of R.O.C. under Grant No. MOST-110-2112-M-001-025,
		the National Natural Science Foundation of China
		under Grant Nos. U1932104, 12142502, 12047503, 12147128, the Guangdong
		Provincial Key Laboratory of Nuclear Science with No. 2019B121203010 and Guangdong Major Project of Basic and Applied Basic Research No. 2020B0301030008.
	\end{acknowledgments}
	
\appendix

\section{Convention and expression of the wave function}
The quark and antiquark fields are expanded in the convention adopted in \cite{Niu:2020}
\begin{equation}
\begin{split}
q(x)&=\int\frac{d\textit{\textbf{p}}}{(2\pi)^{\frac{3}{2}}}\left(\frac{m}{\textit{p}^0}\right)
^{\frac{1}{2}}
\sum_{s}[u_s(\textit{\textbf{p}})b_s(\textit{\textbf{p}})e^{ip\cdot x}+v_s(\textit{\textbf{p}})d_s^{\dagger}(\textit{\textbf{p}})e^{-ip\cdot x}]\;,\\
\bar{q}(x)&=\int\frac{d\textit{\textbf{p}}}{(2\pi)^{\frac{3}{2}}}\left(\frac{m}
{\textit{p}^0}\right)^{\frac{1}{2}}\sum_{s}[\bar{u}_s(\textit{\textbf{p}})b_s^{\dagger}
(\textit{\textbf{p}})e^{-ip\cdot x}+\bar{v}_s(\textit{\textbf{p}})d_s(\textit{\textbf{p}})e^{ip\cdot x}]\;,
\end{split}
\end{equation}
associated with anticommutation relations between the creation and annihilation operators
\begin{align}
\{b_s(\textit{\textbf{p}}),b_{s'}^{\dagger}(\textit{\textbf{p}}')\}=\{d_s(\textit{\textbf{p}}),
d_{s'}^{\dagger}(\textit{\textbf{p}}')\}=\delta_{ss'}\delta^3(\textit{\textbf{p}}-
\textit{\textbf{p}}')\;,
\end{align}
and the normalization relations of spinor
\begin{align}
u_s^{\dagger}(\textit{\textbf{p}})u_{s'}(\textit{\textbf{p}})=v_s^{\dagger}(\textit{\textbf{p}})
v_{s'}(\textit{\textbf{p}})=\left(\frac{p^0}{m}\right)\delta_{ss'}\;.
\end{align}
Then the baryon state in momentum space can be expressed in terms of mock states,
\begin{equation}
\begin{split}
|\B(\textit{\textbf{P}}_c)_{J,M}\rangle&=\sum_{S_z,M_L;c_i}\langle L,M_L;S,S_z|J,M\rangle\int d\textit{\textbf{p}}_1d\textit{\textbf{p}}_2d\textit{\textbf{p}}_3\delta^3(\textit{\textbf{p}}_1
+\textit{\textbf{p}}_2+\textit{\textbf{p}}_3-\textit{\textbf{P}}_c)\Psi_{N,L,M_L}
(\textit{\textbf{p}}_1,\textit{\textbf{p}}_2,\textit{\textbf{p}}_3)\\
&\times \chi_{s_1,s_2,s_3}^{S,S_z}
\frac{\epsilon_{c_1c_2c_3}}{\sqrt{6}}\phi_{i_1,i_2,i_3}b^{\dagger}_{c_1,i_1,s_1,
\textit{\textbf{p}}_1}b^{\dagger}_{c_2,i_2,s_2,\textit{\textbf{p}}_2}b^{\dagger}_{c_3,i_3,s_3,
\textit{\textbf{p}}_3}|0\rangle\;,
\end{split}\label{Baryon}
\end{equation}
which is normalized by
\begin{equation}
\begin{split}
\langle \mathcal{B}(\textit{\textbf{P}}\,'_c )_{J,M}|\mathcal{B}(\textit{\textbf{P}}_c)_{J,M}\rangle=\delta^3(\textit{\textbf{P}}\,'_c-\textit{\textbf{P}}_c)\;.
\end{split}\label{eq:norm}
\end{equation}
In particular, with the quantum numbers defined as
\begin{align}
N=2(n_{\rho}+n_{\lambda})+l_{\rho}+l_{\lambda}, \qquad L=l_{\rho}+l_{\lambda}\;,
\end{align}
the baryon spatial wave function in Eq. (\ref{Baryon}) is
\begin{align}
\Psi_{LM_Ln_{\rho}l_{\rho}n_{\lambda}l_{\lambda}}(\textit{\textbf{P}},\textit{\textbf{p}}_{\rho},\textit{\textbf{p}}_{\lambda})=\delta^3(\textit{\textbf{P}}-\textit{\textbf{P}}_c)\sum_{m}\langle LM_L|l_{\rho }m,l_{\lambda}M_L-m\rangle\psi_{n_{\rho}l_{\rho}m}(\textit{\textbf{p}}_{\rho})\psi_{n_{\lambda}l_{\lambda}(M_L-m)}(\textit{\textbf{p}}_{\lambda})\;,
\label{eq:BaryonWF}
\end{align}
associated with  quark wave function in momentum space
\begin{align}
\psi_{nLm}(\textit{\textbf{p}})=(i)^l(-1)^n\big{[}\frac{2n!}{(n+L+\frac{1}{2})!}\big{]}^\frac{1}{2}\frac{1}{\alpha^{L+\frac{3}{2}}}e^{-\frac{\textit{\textbf{p}}^{2}}{2\alpha^{2}}}L_{n}^{L+\frac{1}{2}}(\frac{\textit{\textbf{p}}^{2}}{\alpha^2})\mathcal{Y}_{Lm}(\textit{\textbf{p}})\;.
\label{wf}
\end{align}
To describe baryon state, the Jacobi coordinate has been introduced in NRQM.
In general, the Jacobi coordinates $\textit{\textbf{x}}_j$ for the $N$-body system are defined as
\begin{equation}
\begin{split}
\textit{\textbf{x}}_j&=\frac{1}{m_{0j}}\sum_{k=1}^{j}m_k\textit{\textbf{r}}_k-\textit{\textbf{r}}_{j+1},\;\{j=1,2...N-1\}\;,\\
\textit{\textbf{x}}_k&=\frac{1}{m_{0N}}\sum_{k=1}^Nm_k\textit{\textbf{r}}_k\;,
\end{split}
\end{equation}
where $m_{0j}=\sum\limits_{k=1}^jm_k$. For the baryon system we have $j=2$ and
the coordinates can be chosen as
\begin{equation}
\begin{split}
&\textit{\textbf{R}}_{c}=\frac{m_1\textit{\textbf{r}}_1+m_2\textit{\textbf{r}}_2+m_3\textit{\textbf{r}}_3}{m_1+m_2+m_3}\;,\\
&\boldsymbol{\rho}=\textit{\textbf{r}}_1-\textit{\textbf{r}}_2\;,\\
&\boldsymbol{\lambda} =\frac{m_1\textit{\textbf{r}}_1+m_2\textit{\textbf{r}}_2}{m_1+m_2}-\textit{\textbf{r}}_3\;.
\end{split}
\end{equation}
By introducing masses of $\rho$- and $\lambda$-mode excitation
$m_{\rho}=\frac{m_1m_2}{m_1+m_2}$ and $m_{\lambda}=\frac{(m_1+m_2)m_3}{(m_1+m_2+m_3)}$, together with baryon mass  $M=m_1+m_2+m_3$, the corresponding
momentums are
%
\begin{equation}
\begin{split}
&\textit{\textbf{p}}=M\dot{\textit{\textbf{R}}_{c}}=\textit{\textbf{p}}_1+\textit{\textbf{p}}_2
+\textit{\textbf{p}}_3\;,\\
&\textit{\textbf{p}}_{\rho}=m_{\rho}\dot{\boldsymbol{\rho}}=\frac{m_2}{m_1+m_2}\textit{\textbf{p}}_1
-\frac{m_1}{m_1+m_2}\textit{\textbf{p}}_2\;,\\
&\textit{\textbf{p}}_{\lambda}=m_{\lambda}\dot{\boldsymbol{\lambda}}=\frac{m_3(\textit{\textbf{p}}_1
+\textit{\textbf{p}}_2)-(m_1+m_2)\textit{\textbf{p}}_3}{(m_1+m_2+m_3)}\;.
\end{split}\label{AppendixA11}
\end{equation}
Then the Hamiltonian to describe a particle interacting in the harmonic oscillator potential,
given by
\begin{equation}
H=\sum_{i=1}^3{ \boldsymbol{p}_i^2\over 2m_i}+{1\over 2}K\sum_{i<j}({\boldsymbol r}_i-{\boldsymbol r}_j)^2
\end{equation}
with  $K$ being a spring constant, becomes the form in terms of Jacobi coordinate
\begin{equation}
H={ {\boldsymbol p}^2\over 2M}+{{\boldsymbol p}_\rho^2\over 2m_\rho} +{{\boldsymbol p}_\lambda^2\over 2m_\lambda}+{1\over 2} m_\rho\omega_\rho^2{\boldsymbol \rho}^2
+{1\over 2} m_\lambda\omega_\lambda^2{\boldsymbol \lambda}^2.
\end{equation}
The harmonic oscillator strengths of the two modes can be further defined as
\begin{align}
\alpha_{\rho}^2=m_{\rho}\omega_{\rho}=\sqrt{\frac{3Km_1m_2}{2(m_1+m_2)}},\qquad \alpha_{\lambda}^2=m_{\lambda}\omega_{\lambda}=\sqrt{\frac{2Km_3(m_1+m_2)}{m_1+m_2+m_3}}\;
\end{align}
for the purpose of convenience.
Hence a useful connection between the two strengths,
\begin{equation}
\alpha_{\lambda}=\left[\frac{4m_3(m_1+m_2)^2}{3m_1m_2(m_1+m_2+m_3)}\right]^{\frac{1}{4}}\alpha_{\rho}\;,
\end{equation}
can be found evidently.  Notice we have an unity Jacobi determinant between the Jacobi coordinate and the ordinary one
in current convention.

In literature, there is another set of convention for Jacobi coordinate, giving
\begin{equation}
\begin{split}
&\tilde{\textit{\textbf{R}}}_{c}=\frac{m_1\textit{\textbf{r}}_1+m_2\textit{\textbf{r}}_2
+m_3\textit{\textbf{r}}_3}{m_1+m_2+m_3}\;,\quad\qquad
\tilde{\textit{\textbf{p}}}=
M\dot{\tilde{\textit{\textbf{R}}_{c}}}=\textit{\textbf{p}}_1+\textit{\textbf{p}}_2
+\textit{\textbf{p}}_3\;,\\
&\tilde{\boldsymbol{\rho}}=\frac{1}{\sqrt{2}}(\textit{\textbf{r}}_1-\textit{\textbf{r}}_2),
\qquad\qquad\qquad\quad\;\;
\tilde{\textit{\textbf{p}}}_{\rho}=\tilde{m}_{\rho}\dot{\tilde{\boldsymbol{\rho}}}
=\sqrt{2}\left(\frac{m_2}{m_1+m_2}\textit{\textbf{p}}_1-\frac{m_1}{m_1+m_2}\textit{\textbf{p}}_2\right)
\;,\\
&\tilde{\boldsymbol{\lambda}} =\sqrt{\frac{2}{3}}\left(\frac{m_1\textit{\textbf{r}}_1+m_2\textit{\textbf{r}}_2}{m_1+m_2}-\textit{\textbf{r}}_3\right)
\;,\quad
\tilde{\textit{\textbf{p}}}_{\lambda}=\tilde{m}_{\lambda}\dot{\tilde{\boldsymbol{\lambda}}}=\sqrt{\frac{3}{2}}\left[\frac{m_3(\textit{\textbf{p}}_1
+\textit{\textbf{p}}_2)-(m_1+m_2)\textit{\textbf{p}}_3}{(m_1+m_2+m_3)}\right]
%
\;,
\end{split}
\label{eq:2ndJacobi}
\end{equation}
with $\rho$- and $\lambda$-type masses
$\tilde{m}_{\rho}=\frac{2m_1m_2}{m_1+m_2}$ and $\tilde{m}_{\lambda}=\frac{3(m_1+m_2)m_3}{2(m_1+m_2+m_3)}$.
Then one can derive the  harmonic oscillator strengths
\begin{equation}
\tilde{\alpha}_{\rho}^2=\tilde{m}_{\rho}\tilde{\omega}_{\rho}=\sqrt{\frac{6Km_1m_2}{(m_1+m_2)}},\qquad
\tilde{\alpha}_{\lambda}^2=\tilde{m}_{\lambda}\tilde{\omega}_{\lambda}=3\sqrt{\frac{Km_3(m_1+m_2)}{2(m_1+m_2+m_3)}}.
\end{equation}
in the tilde convention.
The relations of $\alpha$ parameters between the two conventions,
\begin{equation}
\alpha_{\rho}=\frac{1}{\sqrt{2}}\tilde{\alpha}_{\rho}\;,\qquad
\alpha_{\lambda}=\sqrt{\frac{2}{3}}\tilde{\alpha}_{\lambda}\;,
\label{eq:primeunprime}
\end{equation}
is helpful in the analysis.


\section{Matrix elements of quark operators}
\label{app:ME}
In the pole model calculation of doubly charmed baryon decays, non-perturbative quantities such as  form factors and four-quark operator matrix elements play an essential role.  Here the derivations of these non-perturbative parameters in the  NRQM
are shown in detail in this section.

\subsection{Form factors}

The form factors are defined to parameterize the baryon matrix element of the bilinear quark operator
$\bar{q}'_1\gamma_\mu(1-\gamma_5)q'_2$   have been given in Eq. (\ref{eq:FF}).
For a further
calculation of form factors, we  follow the treatment in Ref. \cite{1989sch} and obtain
 \begin{equation}
\begin{split}
f_1(q^2) &=\langle \mathcal{B}_f(\textit{\textbf{P}}_f)
|\bar{q}'_{1}\gamma_{0}q'_{2}|\mathcal{B}_i(\textit{\textbf{P}}_i)
\rangle\;,\\
g_1(q^2) &=\langle \mathcal{B}_f(\textit{\textbf{P}}_f)
|\bar{q}'_{1}\gamma_{3}\gamma_5q'_{2}|\mathcal{B}_i(\textit{\textbf{P}}_i)
\rangle\;,
\end{split}
\end{equation}
in which  the Breit frame is adopted.  Now by employing the baryon wave functions in
NRQM given by Eq. (\ref{Baryon}), we have
\begin{equation}
\label{eq:AppendixFF}
\begin{split}
f_1(q^2)&=(-1)\times\int d\textit{\textbf{p}}_1d\textit{\textbf{p}}_2d\textit{\textbf{p}}_3d\textit{\textbf{p}}_4d\textit{\textbf{p}}_5d\textit{\textbf{p}}_6d\textit{\textbf{p}}'_{1}d\textit{\textbf{p}}'_{2}\delta^3(\textit{\textbf{p}}_1+\textit{\textbf{p}}_2+\textit{\textbf{p}}_3-\textit{\textbf{P}}_i)\delta^3(\textit{\textbf{p}}_4+\textit{\textbf{p}}_5+\textit{\textbf{p}}_6-\textit{\textbf{P}}_f)\\
&\qquad \times\Psi_f^{\ast}(\textit{\textbf{p}}_4,\textit{\textbf{p}}_5,\textit{\textbf{p}}_6)\Psi_i(\textit{\textbf{p}}_1,\textit{\textbf{p}}_2,\textit{\textbf{p}}_3)\delta^3(\textit{\textbf{p}}'_{1}-\textit{\textbf{p}}'_{2})\langle \mathcal{B}_f\uparrow|{b}_{q'_1}^{\dagger}{b}_{q'_2}|\mathcal{B}_{i}\uparrow\rangle\langle0|b_{6}b_{5}b_{4}{b'}^{\dagger}_{1}{b'}_{2}b^{\dagger}_{1}b^{\dagger}_{2}b^{\dagger}_{3}|0\rangle\\
&=\int d\textit{\textbf{p}}_1d\textit{\textbf{p}}_2d\textit{\textbf{p}}_3d\textit{\textbf{p}}_4d\textit{\textbf{p}}_5d\textit{\textbf{p}}_6d\textit{\textbf{p}}'_{1}d\textit{\textbf{p}}'_{2}\delta^3(\textit{\textbf{p}}_1+\textit{\textbf{p}}_2+\textit{\textbf{p}}_3-\textit{\textbf{P}}_i)\delta^3(\textit{\textbf{p}}_4+\textit{\textbf{p}}_5+\textit{\textbf{p}}_6-\textit{\textbf{P}}_f)\\
&\quad\times\delta^3(\textit{\textbf{p}}_{1}-\textit{\textbf{p}}'_{2})\delta^3(\textit{\textbf{p}}_{4}-\textit{\textbf{p}}'_{1})\delta^3(\textit{\textbf{p}}_{2}-\textit{\textbf{p}}_{6})\delta^3(\textit{\textbf{p}}_{3}-\textit{\textbf{p}}_{5})\delta^3(\textit{\textbf{p}}'_{1}-\textit{\textbf{p}}'_{2})\\
&\;\;\;\;\times\Psi_f^{\ast}(\textit{\textbf{p}}_4,\textit{\textbf{p}}_5,\textit{\textbf{p}}_6)\Psi_i(\textit{\textbf{p}}_1,\textit{\textbf{p}}_2,\textit{\textbf{p}}_3)\langle \mathcal{B}_f\uparrow|{b}_{q'_1}^{\dagger}{b}_{q'_2}|\mathcal{B}_{i}\uparrow\rangle,
\end{split}
\end{equation}
where the baryon wave function defined in Eq. (\ref{eq:BaryonWF}) are denoted as
$\Psi_{i,f}$ concisely. Some details are presented in above two equatoins.
In the first equation, the two matrix elements are calculated in spin-flavor space and
momentum space, respectively, while the factor $-1$ results from a product of
two color wave functions.
Four-momentum $\delta$ functions have been produced
after considering the anti-commutation relations,
corresponding to explicit initial and final state baryons,  in the second equation.
A more compact form of $f_1$
hence can be expressed as
\begin{equation}
 f_1(q^2)=\langle \mathcal{B}_f\uparrow|{b}_{q'_1}^{\dagger}{b}_{q'_2}|\mathcal{B}_{i}\uparrow\rangle X,
 \label{eq:f1}
\end{equation}
with
\begin{equation}
\begin{split}
& X=\int d\textit{\textbf{p}}_1d\textit{\textbf{p}}_2d\textit{\textbf{p}}_3d\textit{\textbf{p}}_4d
\textit{\textbf{p}}_5d\textit{\textbf{p}}_6d\textit{\textbf{p}}'_{1}d\textit{\textbf{p}}'_{2}
\delta^3(\textit{\textbf{p}}_1+\textit{\textbf{p}}_2+\textit{\textbf{p}}_3-\textit{\textbf{P}}_i)
\delta^3(\textit{\textbf{p}}_4+\textit{\textbf{p}}_5+\textit{\textbf{p}}_6-\textit{\textbf{P}}_f)\\
&\quad\times\delta^3(\textit{\textbf{p}}_{1}-\textit{\textbf{p}}'_{2})\delta^3(\textit{\textbf{p}}
_{4}-\textit{\textbf{p}}'_{1})\delta^3(\textit{\textbf{p}}_{2}-\textit{\textbf{p}}_{6})
\delta^3(\textit{\textbf{p}}_{3}-\textit{\textbf{p}}_{5})\delta^3(\textit{\textbf{p}}'_{1}
-\textit{\textbf{p}}'_{2})\Psi_f^{\ast}(\textit{\textbf{p}}_4,\textit{\textbf{p}}_5,
\textit{\textbf{p}}_6)\Psi_i(\textit{\textbf{p}}_1,\textit{\textbf{p}}_2,\textit{\textbf{p}}_3)\;.
\end{split}
\end{equation}
To be specific, the convention for matrix element in Eq. (\ref{eq:f1}) in spin-flavor
space keeps the same as our previous work \cite{Cheng:2020wmk}.
A similar derivation leads to
\begin{equation}
 g_1(q^2)=\langle \mathcal{B}_f\uparrow|{b}_{q'_1}^{\dagger}{b}_{q'_2}\sigma_z
 |\mathcal{B}_{i}\uparrow\rangle X,
 \label{eq:g1}
\end{equation}
in which a common spatial wave function integral $X$ has been shared with the form factor
$f_1(q^2)$.
As for the axial-vector form factor $g_{\mathcal{B}'\mathcal{B}}^{A(P)}$,
it differs $g_1$ from its spatial wave function integral $Y$, giving
\begin{equation}
 g_{\mathcal{B}'\mathcal{B}}^{A(P)}(q^2)
=\langle \mathcal{B}_i\uparrow|{b}_{q'_1}^{\dagger}{b}_{q'_2}\sigma_z
|\mathcal{B}_{f}\uparrow\rangle Y, \\
\end{equation}
with
\begin{equation}
\begin{split}
& Y=\int d\textit{\textbf{p}}_1d\textit{\textbf{p}}_2d\textit{\textbf{p}}_3d\textit{\textbf{p}}_4d
\textit{\textbf{p}}_5d\textit{\textbf{p}}_6d\textit{\textbf{p}}'_{1}d\textit{\textbf{p}}'_{2}
\delta^3(\textit{\textbf{p}}_1+\textit{\textbf{p}}_2+\textit{\textbf{p}}_3-\textit{\textbf{P}}_i)
\delta^3(\textit{\textbf{p}}_4+\textit{\textbf{p}}_5+\textit{\textbf{p}}_6-\textit{\textbf{P}}_f)\\
&\quad \times\delta^3(\textit{\textbf{p}}_{3}-\textit{\textbf{p}}'_{2})\delta^3(\textit{\textbf{p}}_{6}
-\textit{\textbf{p}}'_{1})\delta^3(\textit{\textbf{p}}_{1}-\textit{\textbf{p}}_{4})\delta^3
(\textit{\textbf{p}}_{2}-\textit{\textbf{p}}_{5})\delta^3(\textit{\textbf{p}}'_{1}
-\textit{\textbf{p}}'_{2})\Psi_f^{\ast}(\textit{\textbf{p}}_4,\textit{\textbf{p}}_5,
\textit{\textbf{p}}_6)\Psi_i(\textit{\textbf{p}}_1,\textit{\textbf{p}}_2,\textit{\textbf{p}}_3).
\end{split}
\label{eq:gA}
\end{equation}
Though the two integrals $X$ and $Y$ formally share the same structure, their
difference exists in the wave functions of two baryons and hence
brings a further difference in $\delta$ functions.  In fact,  $f_1$ and $g_1$
depict the transition between the initial and final state baryons while
$g_{\mathcal{B}'\mathcal{B}}^{A(P)}$ is devoted to the one between initial
and intermediate (or intermediate and final) baryons.
 Therefore such a difference between $X$ and $Y$ is a consequence of the pole model.

\subsection{Matrix elements of four-quark operators}
We continue woking in NR quark model to calculate the  matrix element of
four-quark operator $(\bar{q'}_{1}{q'}_{2})(\bar{q'}_{3}{q'}_{4})$. With the help of
baryon wave function Eq. (\ref{Baryon}), it can be expanded as
\begin{equation}
\begin{split}
&\langle \mathcal{B}_f(\textit{\textbf{P}}_f)
|(\bar{q'}_{1}{q'}_{2})(\bar{q'}_{3}{q'}_{4})|\mathcal{B}_i(\textit{\textbf{P}}_i)
\rangle\\
=&\int d\textit{\textbf{p}}_1d\textit{\textbf{p}}_2d\textit{\textbf{p}}_3d\textit{\textbf{p}}_4d
\textit{\textbf{p}}_5d\textit{\textbf{p}}_6d\textit{\textbf{p}}'_{1}d\textit{\textbf{p}}'_{2}d
\textit{\textbf{p}}'_{3}d\textit{\textbf{p}}'_{4}\delta^3(\textit{\textbf{p}}_1+
\textit{\textbf{p}}_2+\textit{\textbf{p}}_3-\textit{\textbf{P}}_i)\delta^3(\textit{\textbf{p}}_4
+\textit{\textbf{p}}_5+\textit{\textbf{p}}_6-\textit{\textbf{P}}_f)\\
&\quad\times\Psi_f^{\ast}(\textit{\textbf{p}}_4,\textit{\textbf{p}}_5,\textit{\textbf{p}}_6)
\Psi_i(\textit{\textbf{p}}_1,\textit{\textbf{p}}_2,\textit{\textbf{p}}_3)\delta^3
(\textit{\textbf{p}}'_{1}+\textit{\textbf{p}}'_{3}-\textit{\textbf{p}}'_{2}-\textit
{\textbf{p}}'_{4})\\
&\quad\times\langle \mathcal{B}_f\uparrow|
({b}^{\dagger}_{q'_1}{b}_{q'_2})_1({b}^{\dagger}_{q'_3}{b}_{q'_4})_2
(1-\boldsymbol{\sigma}_1\cdot\boldsymbol{\sigma}_2)|\mathcal{B}_i\uparrow\rangle\langle0|b_{6}
b_{5}b_{4}{b'}^{\dagger}_{1}{b'}_{2}{b'}^{\dagger}_{3}{b'}_{4}b^{\dagger}_{1}b^{\dagger}_{2}
b^{\dagger}_{3}|0\rangle\\
=&\int d\textit{\textbf{p}}_1d\textit{\textbf{p}}_2d\textit{\textbf{p}}_3d\textit{\textbf{p}}_4d
\textit{\textbf{p}}_5d\textit{\textbf{p}}_6d\textit{\textbf{p}}'_{1}d\textit{\textbf{p}}'_{2}d
\textit{\textbf{p}}'_{3}d\textit{\textbf{p}}'_{4}\delta^3(\textit{\textbf{p}}_1+\textit
{\textbf{p}}_2+\textit{\textbf{p}}_3-\textit{\textbf{P}}_i)\delta^3(\textit{\textbf{p}}_4
+\textit{\textbf{p}}_5+\textit{\textbf{p}}_6-\textit{\textbf{P}}_f)\\
&\quad\times\Psi_f^{\ast}(\textit{\textbf{p}}_4,\textit{\textbf{p}}_5,\textit{\textbf{p}}_6)
\Psi_i(\textit{\textbf{p}}_1,\textit{\textbf{p}}_2,\textit{\textbf{p}}_3)\delta^3(\textit
{\textbf{p}}'_{1}+\textit{\textbf{p}}'_{3}-\textit{\textbf{p}}'_{2}-\textit{\textbf{p}}'_{4})
\delta^3(\textit{\textbf{p}}'_{2}-\textit{\textbf{p}}_{1})\delta^3(\textit{\textbf{p}}'_{4}
-\textit{\textbf{p}}_{3})\\
&\quad\times\delta^3(\textit{\textbf{p}}_{4}-\textit{\textbf{p}}'_{1})\delta^3(\textit
{\textbf{p}}_{5}-\textit{\textbf{p}}'_{3})\delta^3(\textit{\textbf{p}}_{6}-\textit{\textbf{p}}_{2})
\langle \mathcal{B}_f\uparrow|
({b}^{\dagger}_{q'_1}{b}_{q'_2})_1({b}^{\dagger}_{q'_3}{b}_{q'_4})_2
(1-\boldsymbol{\sigma}_1\cdot\boldsymbol{\sigma}_2)|\mathcal{B}_i\uparrow\rangle
\end{split}\label{eq:4q}
\end{equation}
Two equations are presented with the similar treatment to the form factor in  Eq. (\ref{eq:f1}),
then a compact form can be achieved as
\begin{equation}
\langle \mathcal{B}_f(\textit{\textbf{P}}_f)
|(\bar{q'}_{1}{q'}_{2})(\bar{q'}_{3}{q'}_{4})|\mathcal{B}_i(\textit{\textbf{P}}_i)
\rangle
=
\langle \mathcal{B}_f\uparrow|
({b}^{\dagger}_{q'_1}{b}_{q'_2})_1({b}^{\dagger}_{q'_3}{b}_{q'_4})_2
(1-\boldsymbol{\sigma}_1\cdot\boldsymbol{\sigma}_2)|\mathcal{B}_i\uparrow\rangle Z
\end{equation}
with
\begin{equation}
\begin{split}
Z&=\int d\textit{\textbf{p}}_1d\textit{\textbf{p}}_2d\textit{\textbf{p}}_3d\textit{\textbf{p}}_4d
\textit{\textbf{p}}_5d\textit{\textbf{p}}_6d\textit{\textbf{p}}'_{1}d\textit{\textbf{p}}'_{2}d
\textit{\textbf{p}}'_{3}d\textit{\textbf{p}}'_{4}\delta^3(\textit{\textbf{p}}_1+\textit
{\textbf{p}}_2+\textit{\textbf{p}}_3-\textit{\textbf{P}}_i)\delta^3(\textit{\textbf{p}}_4
+\textit{\textbf{p}}_5+\textit{\textbf{p}}_6-\textit{\textbf{P}}_f)\\
&\;\;\;\;\times\delta^3(\textit{\textbf{p}}'_{2}-\textit{\textbf{p}}_{1})\delta^3(\textit
{\textbf{p}}'_{4}-\textit{\textbf{p}}_{3})\delta^3(\textit{\textbf{p}}_{4}-\textit{\textbf{p}}'_{1})
\delta^3(\textit{\textbf{p}}_{5}-\textit{\textbf{p}}'_{3})\delta^3(\textit{\textbf{p}}_{6}
-\textit{\textbf{p}}_{2})\delta^3(\textit{\textbf{p}}'_{1}+\textit{\textbf{p}}'_{3}
-\textit{\textbf{p}}'_{2}-\textit{\textbf{p}}'_{4})\\
&\;\;\;\;\times\Psi_f^{\ast}(\textit{\textbf{p}}_4,\textit{\textbf{p}}_5,\textit{\textbf{p}}_6)
\Psi_i(\textit{\textbf{p}}_1,\textit{\textbf{p}}_2,\textit{\textbf{p}}_3)\;.
\end{split}\label{eq:4quark}
\end{equation}
Here the spatial integral $Z$ is taken between initial (final) and intermediate baryons with four quark fields involved, which brings in one more $\delta$ function compared with the case
in integral $Y$.

\subsection{Momentum integrals of baryon wave functions}
After integrating all the $\delta$ functions in Eqs. (\ref{eq:f1}), (\ref{eq:gA}) and (\ref{eq:4q}),
keeping the momentum conservation, we have  concise expressions of $X, Y, Z$ in terms of
momentum integrals of baryon wave functions,
\begin{equation}
\begin{split}
& X=\delta^3(\textit{\textbf{P}}_i-\textit{\textbf{P}}_f)\int d\textit{\textbf{p}}_2d\textit{\textbf{p}}_3\Psi_f^{\ast}((\textit{\textbf{P}}_i-\textit{\textbf{p}}_2-\textit{\textbf{p}}_3),\textit{\textbf{p}}_3,\textit{\textbf{p}}_2)\Psi_i((\textit{\textbf{P}}_i-\textit{\textbf{p}}_2-\textit{\textbf{p}}_3),\textit{\textbf{p}}_2,\textit{\textbf{p}}_3)\;,\\
&Y=\delta^3(\textit{\textbf{P}}_i-\textit{\textbf{P}}_f)\int d\textit{\textbf{p}}_2d\textit{\textbf{p}}_3\Psi_f^{\ast}((\textit{\textbf{P}}_i-\textit{\textbf{p}}_2-\textit{\textbf{p}}_3),\textit{\textbf{p}}_2,\textit{\textbf{p}}_3)\Psi_i((\textit{\textbf{P}}_i-\textit{\textbf{p}}_2-\textit{\textbf{p}}_3),\textit{\textbf{p}}_2,\textit{\textbf{p}}_3)\;,\\
&Z=\delta^3(\textit{\textbf{P}}_i-\textit{\textbf{P}}_f)\int d\textit{\textbf{p}}_1d\textit{\textbf{p}}_2d\textit{\textbf{p}}_4\Psi_f^{\ast}(\textit{\textbf{p}}_4,(\textit{\textbf{P}}_i-\textit{\textbf{p}}_2-\textit{\textbf{p}}_4),\textit{\textbf{p}}_2)\Psi_i(\textit{\textbf{p}}_1,\textit{\textbf{p}}_2,(\textit{\textbf{P}}_i-\textit{\textbf{p}}_1-\textit{\textbf{p}}_2))\;.\\
\end{split}
\label{eq:X,Y,Z}
\end{equation}
The integrals $X$ and $Y$ are similar except the interchange of ${\bf p}_2$ and ${\bf p}_3$ in the wave function $\Psi_f$.
Then the remaining task is to evaluate them.
Before proceeding to a detailed calculation, it is useful to firstly deal with
the product of two wave functions as it is
the common part in all the three integrals. Taking the one in $X$ as an example,
a direct calculation leads to
\begin{equation}
\Psi^{\ast}_f(\textit{\textbf{p}}_4,\textit{\textbf{p}}_5,\textit{\textbf{p}}_6)\Psi_i(\textit{\textbf{p}}_1,\textit{\textbf{p}}_2,\textit{\textbf{p}}_3)
=
\frac{1}{\pi^{3}}
\left(\alpha_{\rho 1}\alpha_{\lambda 1}\alpha_{\rho 2}\alpha_{\lambda 2}\right)^{-\frac{3}{2}}
e^{-\frac{1}{2}
\left(\frac{\textit{\textbf{p}}_{\rho i}^2}{\alpha_{\rho 1}^2}+\frac{\textit{\textbf{p}}_{\lambda i}^2}{\alpha_{\lambda 1}^2}+\frac{\textit{\textbf{p}}_{\rho f}^2}{\alpha_{\rho 2}^2}+\frac{\textit{\textbf{p}}_{\lambda f}^2}{\alpha_{\lambda 2}^2}\right)},
\label{eq:WFproduct}
\end{equation}
which is based on the relations between two coordinates
\begin{equation}
\begin{split}
&\textit{\textbf{p}}_{\rho i}=\frac{m_2}{m_1+m_2}\textit{\textbf{p}}_1-\frac{m_1}{m_1+m_2}\textit{\textbf{p}}_2\;,\qquad\textit{\textbf{p}}_{\rho f}=\frac{m_5}{m_4+m_5}\textit{\textbf{p}}_4-\frac{m_4}{m_4+m_5}\textit{\textbf{p}}_5\;,\\
&\textit{\textbf{p}}_{\lambda i}=\frac{m_3(\textit{\textbf{p}}_1+\textit{\textbf{p}}_2)-(m_1+m_2)\textit{\textbf{p}}_3}{(m_1+m_2+m_3)}\;,\quad\textit{\textbf{p}}_{\lambda f}=\frac{m_6(\textit{\textbf{p}}_4+\textit{\textbf{p}}_5)-(m_4+m_5)\textit{\textbf{p}}_6}{(m_4+m_5+m_6)}\;.
\end{split}\label{eq:p-Jacb}
\end{equation}
A replacement of the index $2\to 3$ in Eq. (\ref{eq:WFproduct})
gives the one in $Y$ while $1\to 3$
provides the corresponding one to $Z$.
Taking a static limit of initial baryon $\textit{\textbf{P}}_{i}=0$,
a further calculation for $X$ yields
\begin{equation}
X=\frac{1}{\pi^{3}}
\left(\alpha_{\rho 1}\alpha_{\lambda 1}\alpha_{\rho 2}\alpha_{\lambda 2}\right)^{-\frac{3}{2}}
\left(\frac{4\pi^2}{a_Xb_X-\frac{c^2_X}{4}}\right)^{3/2}=d_X
\left(\alpha_{\rho 1}\alpha_{\lambda 1}\alpha_{\rho 2}\alpha_{\lambda 2}\right)^{-\frac{3}{2}}\;,
\label{eq:X}
\end{equation}
associated with the auxiliary parameters, giving
\begin{eqnarray}
d_X=8\left(a_Xb_X-\frac{c^2_X}{4}\right)^{-3/2},
\end{eqnarray}
with
\begin{eqnarray}
  a_X =\frac{1}{\alpha_{\rho 1}^2}+\frac{1}{\alpha_{\lambda 2}^2}+\frac{m_u^2}{(m_s+m_u)^2\alpha_{\rho 2}^2},\quad
b_X=\frac{1}{\alpha_{\rho 2}^2}+\frac{1}{\alpha_{\lambda 1}^2}+\frac{1}{4\alpha_{\rho 1}^2},\quad
c_X=\frac{1}{\alpha_{\rho 1}^2}+\frac{2m_u}{(m_s+m_u)\alpha_{\rho 2}^2}\;.\nonumber
\end{eqnarray}
Likewise, $Y$ and $Z$ can be derived as
\begin{eqnarray}
Y  = d_Y
\left(\alpha_{\rho 1}\alpha_{\lambda 1}\alpha_{\rho 3}\alpha_{\lambda 3}\right)^{-\frac{3}{2}}\;,
\qquad
Z  = d_Z
\left(
{\alpha_{\rho 3}\alpha_{\lambda 3} \alpha_{\rho 2}
\alpha_{\lambda 2}}\right)^{-\frac{3}{2}}
\;,
\label{eq:YZ}
\end{eqnarray}
together with
\begin{equation}
\begin{split}
d_Y&=8\left(a_Yb_Y-\frac{c^2_Y}{4}\right)^{-3/2}\;,
\qquad\qquad\qquad\qquad
d_Z=8 \left(2\pi \alpha_{\rho 2}^2\right)^{3/2}
\left(a_Zb_Z-\frac{c^2_Z}{4}\right)^{-3/2}\;,
\\
a_Y&=\frac{1}{\alpha_{\rho 1}^2}+\frac{1}{\alpha_{\rho 3}^2}\;,
\qquad\qquad \qquad\qquad\qquad\qquad\;
a_Z=\frac{1}{4\alpha_{\rho 3}^2}+\frac{1}{\alpha_{\lambda 3}^2}\;,\\
b_Y&=\frac{1}{\alpha_{\lambda 3}^2}+\frac{1}{\alpha_{\lambda 1}^2}
+\frac{1}{4\alpha_{\rho 1}^2}+\frac{1}{4\alpha_{\rho 3}^2}\;,
\qquad\qquad\quad\;
b_Z=\frac{1}{4\alpha_{\rho 3}^2}+\frac{1}{\alpha_{\lambda 3}^2}+\frac{1}{\alpha_{\lambda 2}^2}\;,\\
c_Y&=\frac{1}{\alpha_{\rho 1}^2}+\frac{1}{\alpha_{\rho 3}^2}\;,
\qquad\qquad\qquad\quad\qquad\qquad\;\;\;\;\;
c_Z=-2\left(\frac{1}{4\alpha_{\rho 3}^2}-\frac{1}{\alpha_{\lambda 3}^2}\right)
\;.\\
\end{split}\label{eq:AppendixB3}
\end{equation}
Eqs. (\ref{eq:X}) and (\ref{eq:YZ}) give the final expressions
of the momentum integrals.


\end{document}